\documentclass[11pt,a4paper]{article}

\usepackage[utf8]{inputenc}
\usepackage[T1]{fontenc}
\usepackage{amsmath,amssymb,amsfonts}
\usepackage{graphicx}
\usepackage{booktabs}
\usepackage{hyperref}
\usepackage{algorithm}
\usepackage{algorithmic}
\usepackage{multirow}
\usepackage{xcolor}
\usepackage{enumitem}
\usepackage{subcaption}
\usepackage[margin=1in]{geometry}
\usepackage{float}
\usepackage{natbib}

\graphicspath{{figures/}}

\newcommand{\Dreal}{\mathcal{D}_{\text{real}}}
\newcommand{\Dsynth}{\mathcal{D}_{\text{synth}}}
\newcommand{\Schema}{\mathcal{S}}

\title{%
  Beyond Marginals: A Multi-Dimensional Evaluation Framework \\
  for Multi-Table Synthetic Data Generation%
}

\author{%
  \begin{tabular}[t]{c}
    Aparana Gupta \\
    \texttt{aparanagupta@microsoft.com}
  \end{tabular}
  \hspace{1.0em}
  \begin{tabular}[t]{c}
    Anurup Dey \\
    \texttt{anurupdey@microsoft.com}
  \end{tabular}
  \hspace{1.0em}
  \begin{tabular}[t]{c}
    Suyash Dwivedi \\
    \texttt{sudwivedi@microsoft.com}
  \end{tabular}
}

\date{\today}

\begin{document}

\maketitle

\begin{abstract}
Synthetic data generation is critical for privacy compliance, machine
learning augmentation, and software testing. While single-table
evaluation is well-established, multi-table (relational) synthesis---the
dominant enterprise use case---lacks a unified evaluation framework.
Existing approaches assess marginal column distributions in isolation,
overlooking joint distributions, cross-table structural integrity,
downstream utility, and production-readiness edge cases.

We present \textbf{SynEval}, a six-dimensional evaluation framework for
multi-table synthetic databases. SynEval jointly assesses per-column
fidelity, multivariate structure preservation (including a novel
conditional distribution check), cross-table integrity, ML utility,
privacy protection, and edge-case robustness. The framework produces a
unified weighted quality score with per-table, per-dimension
drill-down, and is generator-agnostic---operating on any pair of real
and synthetic CSV folders with automatic schema inference.

To our knowledge, SynEval is the first framework to combine conditional
distribution checks $P(Y \mid X)$, cross-table cardinality validation,
and production-readiness edge-case testing within a single evaluation
pipeline for relational synthetic data.
\end{abstract}

\section{Introduction}
\label{sec:intro}

Synthetic data generation has progressed from a niche academic pursuit
to a foundational enterprise technology, driven by privacy regulations
such as GDPR and HIPAA~\citep{eu_gdpr_2016}, ML data
augmentation~\citep{jordon2022synthetic}, and software
testing~\citep{lu2023machine, el2020practical}. In practice, enterprise
data resides in multi-table relational databases connected by
primary--foreign key relationships, making multi-table synthesis the
production-relevant problem. While generation methods have advanced
rapidly~\citep{patki2016sdv, xu2019modeling, solatorio2023realtabformer,
pang2024clavaddpm}, the complementary problem of multi-table
\emph{evaluation} has received far less systematic attention. The current
state of practice is characterized by several critical gaps:

\begin{enumerate}[leftmargin=*]
  \item \textbf{Marginal-only evaluation.} Standard approaches evaluate
  each column independently via KS or JSD, ignoring whether inter-column
  correlations and conditional distributions are preserved.

  \item \textbf{Cross-table integrity ignored.} Most frameworks treat
  tables as isolated entities, neglecting FK referential integrity,
  cardinality distributions, and cross-table statistical correlations.

  \item \textbf{Fragmented utility and privacy assessment.} ML utility
  and privacy metrics are evaluated in standalone analyses, not
  integrated into a holistic score enabling cross-generator comparison.

  \item \textbf{Missing production-readiness checks.} Real-world data
  exhibits null patterns, rare categories, zero-inflated columns, and
  outlier tails that are invisible to standard distribution tests.

  \item \textbf{No unified scoring.} Without a common scoring system,
  comparing generators requires ad hoc, non-reproducible analysis.
\end{enumerate}

\paragraph{Our contributions.}
We address these gaps by proposing \textbf{SynEval}, a comprehensive,
six-dimensional evaluation framework for multi-table synthetic data:

\begin{itemize}[leftmargin=*]
  \item \textbf{Formalization of six evaluation dimensions} with
  rigorous mathematical definitions: per-column fidelity, multivariate
  structure, cross-table integrity, ML utility, privacy, and
  edge-case robustness.

  \item \textbf{Novel metrics} not present in prior frameworks:
  conditional distribution checks $P(Y \mid X \in \text{bin})$,
  cross-table cardinality distribution matching, and a comprehensive
  edge-case test battery covering null patterns, rare categories,
  zero-inflation, tail behavior, mode collapse, and percentile accuracy.

  \item \textbf{A unified weighted scoring system} with
  per-table, per-dimension drill-down---enabling direct, reproducible
  comparison across generators.

  \item \textbf{A generator-agnostic, open-source implementation} with
  automatic schema inference and a publication-quality visualization
  suite.
\end{itemize}

\section{Related Work}
\label{sec:related}

We organize related work into four areas: single-table evaluation,
multi-table generators, multi-table evaluation, and privacy-specific
evaluation.

\subsection{Single-Table Synthetic Data Evaluation}
\label{sec:related:single}

The \textbf{SDMetrics} library~\citep{patki2016sdv} provides
column-level quality metrics including KS complementary score, TVD
complementary score, and contingency similarity, aggregated into a
single quality score per table. While SDMetrics extends to multi-table
evaluation with cardinality shape similarity and parent-child
detection (see \S\ref{sec:related:multitable}), its per-table metrics
remain marginal-only. \textbf{TableEvaluator}~\citep{bowen2020comparative}
offers KS and chi-square tests for comparing real and synthetic
distributions.
\citet{chundawat2022tabsyndex} propose TabSynDex, a composite index
combining fidelity, utility, and privacy for single tables.
TableEvaluator and TabSynDex treat each table independently and
provide no assessment of relational structure preservation.
\citet{dankar2022multi} and \citet{wang2023synthetic}
highlight the multi-objective nature of synthetic data quality but
remain at the single-table level.

\subsection{Multi-Table Synthetic Data Generators}
\label{sec:related:generators}

Multi-table generation is an active area. \textbf{SDV}~\citep{patki2016sdv}
introduced the Hierarchical Modeling Algorithm (HMA), fitting Gaussian
copula models per table and sampling in parent-to-child order.
\textbf{RC-TGAN}~\citep{gueye2023rctgan} extends conditional GAN
training to relational data by incorporating parent-row features into
the child table generator, with grandparent propagation for deeper
hierarchies. \textbf{ClavaDDPM}~\citep{pang2024clavaddpm} uses
classifier-guided diffusion models with Gaussian Mixture clustering as
latent intermediaries between parent and child tables, extending to
multi-parent scenarios through majority voting. Each generator evaluates
its output using different metrics---a primary motivation for a
standardized evaluation framework.

\subsection{Multi-Table Evaluation}
\label{sec:related:multitable}

Systematic evaluation frameworks for multi-table synthetic data remain
sparse, in contrast to the single-table ecosystem.

\textbf{SDMetrics}~\citep{patki2016sdv} is the only widely-used package
supporting multi-table evaluation. It extends single-table metrics with
cardinality shape similarity (comparing child-count distributions per
parent via KS test) and parent-child detection using logistic
regression. However, it does not assess conditional distributions within
subgroups, cross-table aggregate correlations, or production-readiness
edge cases. Furthermore, SDMetrics aggregates all results into a single
score, masking per-table and per-dimension diagnostic information.

\citet{dankar2022multi} propose a multi-dimensional evaluation along
four axes---attribute fidelity, bivariate fidelity, population fidelity,
and application fidelity---but their evaluation operates exclusively on
single tables with no relational awareness.

\textbf{SynthEval}~\citep{lautrup2025syntheval} provides a detailed
utility and privacy evaluation framework for tabular synthetic data,
combining statistical and ML-based techniques. However, it is limited
to single-table evaluation and does not address cross-table integrity.

\citet{lautrup2024systematic} present a systematic review of generative
modelling tools and utility metrics for synthetic tabular data,
cataloguing fidelity, privacy, and utility measures across tools
including SDMetrics, SynthCity, and Table Evaluator. Their survey
confirms that multi-table evaluation remains an open gap.

\citet{hudovernik2024syntherela} (SyntheRela) make the most directly
relevant contribution, introducing \emph{Discriminative Detection}
(DD)---a classification-based fidelity test using XGBoost---and extending
it to multi-table fidelity via aggregation (DDA). They benchmark six
methods across six datasets and show that detection-based approaches
with aggregated child-table features outperform logistic detection and
traditional metrics at identifying synthetic data. However, SyntheRela
focuses on detection-based fidelity and ML utility, but does not
evaluate conditional distribution preservation $P(Y \mid X)$, privacy
metrics (DCR, NNDR), or production-readiness edge cases (null patterns,
rare categories, zero-inflation, bounds compliance).

\citet{ward2026mtmia} address multi-table privacy specifically,
proposing MT-MIA (Multi-Table Membership Inference Attack) and
demonstrating that relational structure creates additional privacy
vulnerabilities not captured by single-table metrics.

To our knowledge, no existing framework jointly evaluates all six
dimensions we propose---per-column fidelity, multivariate structure
(including conditional distributions), cross-table integrity, ML
utility, privacy, and edge-case robustness---within a single pipeline
that produces a unified quality score.

\subsection{Privacy Evaluation for Generative Models}
\label{sec:related:privacy}

Privacy evaluation has established several key metrics: \textbf{Distance
to Closest Record (DCR)}~\citep{yale2019assessing} measures Euclidean
distance from synthetic to nearest real record in standardized feature
space; \textbf{NNDR}~\citep{yale2020generation} computes the ratio of
first-to-second nearest neighbor distances; and \textbf{membership
inference attacks}~\citep{shokri2017membership} test whether an
adversary can identify training records. These are well-established
individually; our contribution is their integration into a composite
privacy dimension weighted alongside fidelity, utility, and structural
metrics.

\section{Problem Formulation}
\label{sec:formulation}

\subsection{Setting and Notation}
\label{sec:formulation:notation}

Let $\Dreal = \{T_1, T_2, \ldots, T_K\}$ denote a real relational
database comprising $K$ tables, with schema $\Schema$ specifying:

\begin{itemize}[leftmargin=*]
  \item $\text{PK}(T_k)$: the primary key column of table $T_k$.
  \item $\text{FK}(T_k)$: a set of foreign key references
        $\{(c, T_p, c') \mid c \in T_k \text{ references } c' \in T_p\}$.
  \item $\text{Rel} = \{(T_p, c_p, T_c, c_c)\}$: the set of
        parent--child relationships in the database.
  \item Column type annotations: for each table $T_k$, a partition of
        non-key columns into continuous columns $\mathcal{C}_k^{\text{cont}}$
        and categorical columns $\mathcal{C}_k^{\text{cat}}$.
\end{itemize}

A \emph{multi-table synthetic data generator} is a function
$G: (\Dreal, \Schema) \longrightarrow \Dsynth$
that produces a synthetic database $\Dsynth = \{\hat{T}_1, \hat{T}_2,
\ldots, \hat{T}_K\}$ with the same schema $\Schema$.

\subsection{The Quality Function}
\label{sec:formulation:quality}

Our objective is to define a quality function
$Q(\Dreal, \Dsynth, \Schema) \in [0, 1]$
satisfying three properties:

\begin{enumerate}[leftmargin=*]
  \item \textbf{Comprehensiveness.} $Q$ captures statistical fidelity,
  structural integrity, downstream utility, privacy protection, and
  production robustness---not merely marginal distributions.

  \item \textbf{Decomposability.} $Q$ decomposes into per-table,
  per-dimension scores with row-count weighting:
  \begin{equation}
    Q = \frac{\sum_{k=1}^{K} n_k \cdot Q_k}{\sum_{k=1}^{K} n_k}, \qquad
    Q_k = \sum_{d=1}^{D} w_d \cdot s_{k,d}
  \end{equation}
  where $n_k = |T_k|$ is the row count of table $T_k$,
  $s_{k,d} \in [0, 1]$ is the score of table $T_k$ on
  dimension $d$, and $w_d$ are configurable weights with
  $\sum_d w_d = 1$. Row-count weighting ensures that large
  transaction tables (e.g., 420k rows) dominate the global score
  over small lookup tables (e.g., 120 rows), reflecting their
  greater practical importance.

  \item \textbf{Generator-agnosticism.} $Q$ operates on the output
  CSVs only, with no coupling to generator internals.
\end{enumerate}

\subsection{Failure Mode Taxonomy}
\label{sec:formulation:failures}

We identify nine distinct failure modes that a comprehensive evaluation
must detect (Table~\ref{tab:failure_modes}). Crucially, several of
these (modes 2--4, 8--9) are invisible to standard per-column distribution
tests, motivating a multi-dimensional framework.

\begin{table}[H]
\centering
\caption{Taxonomy of multi-table synthetic data failure modes.}
\label{tab:failure_modes}
\small
\begin{tabular}{@{}p{3.5cm}p{5.5cm}p{3.5cm}@{}}
\toprule
\textbf{Failure Mode} & \textbf{Example} & \textbf{Detecting Dimension} \\
\midrule
Marginal drift &
\texttt{salary} distribution shifted by 20\% &
Per-Column Fidelity \\
\addlinespace
Correlation destruction &
\texttt{hours} $\times$ \texttt{billing\_rate} correlation lost &
Multivariate Structure \\
\addlinespace
Conditional collapse &
$P(\text{salary} \mid \text{dept}=\text{Eng.})$ is flat &
Multivariate Structure \\
\addlinespace
Orphaned foreign keys &
Child row references non-existent parent PK &
Cross-Table Integrity \\
\addlinespace
Cardinality distortion &
Real: 5--50 txns/account; Synth: always 25 &
Cross-Table Integrity \\
\addlinespace
ML degradation &
Model on synth achieves 60\% of real AUC &
ML Utility \\
\addlinespace
Record leakage &
Synthetic rows are near-copies of real records &
Privacy \\
\addlinespace
Membership inference &
A subject's presence in the training data is detectable via
categorical presence or ML confidence &
Privacy \\
\addlinespace
Edge-case erasure &
Null patterns, rare categories lost &
Edge Cases \\
\bottomrule
\end{tabular}
\end{table}

\section{SynEval: Proposed Framework}
\label{sec:framework}

\subsection{Architecture Overview}
\label{sec:framework:overview}

\begin{figure}[H]
\centering
\includegraphics[width=0.95\textwidth]{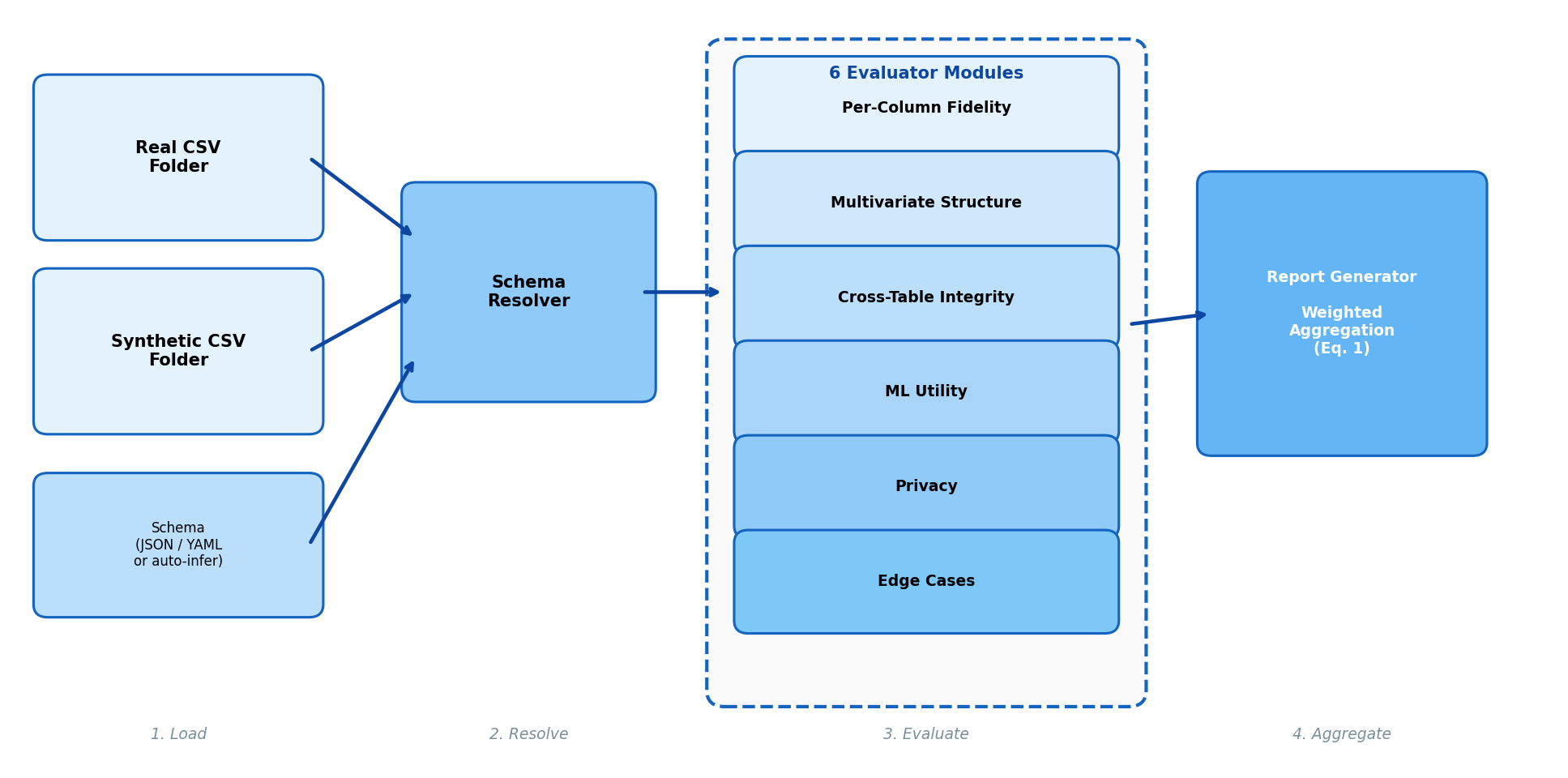}
\caption{SynEval pipeline architecture.}
\label{fig:architecture}
\end{figure}

SynEval operates as a five-stage pipeline
(Figure~\ref{fig:architecture}): (1)~load real and synthetic CSV
folders, matching tables by filename; (2)~resolve schema from YAML or
auto-infer column types (see Appendix~\ref{app:details:schema});
(3)~evaluate per-table dimensions (Fidelity, Multivariate, ML Utility,
Privacy, Edge Cases); (4)~evaluate global Cross-Table Integrity; and
(5)~aggregate scores via configurable weights.

The default dimension weights (Table~\ref{tab:weights}) are fully
configurable.

\begin{table}[H]
\centering
\caption{Dimension weights used in the comparison study.}
\label{tab:weights}
\small
\begin{tabular}{@{}lcc@{}}
\toprule
\textbf{Dimension} & \textbf{Weight ($w_d$)} & \textbf{Rationale} \\
\midrule
Per-Column Fidelity     & 0.25 & Baseline statistical correctness \\
Multivariate Structure  & 0.20 & Joint distribution preservation \\
Cross-Table Integrity   & 0.15 & Relational structure fidelity \\
ML Utility              & 0.20 & Downstream task fitness \\
Privacy                 & 0.10 & Re-identification risk \\
Edge Cases              & 0.10 & Production robustness \\
\bottomrule
\end{tabular}
\end{table}

\paragraph{ML Utility sensitivity.}
The ML Utility dimension (TSTR efficacy ratio) depends on target column
selection and is therefore more variable across tables than purely
statistical dimensions. It is nonetheless included in the composite
score to ensure that downstream task fitness is reflected in the
overall quality assessment. Users may override the default target via
configuration when domain knowledge is available.

\subsection{Dimension 1: Per-Column Fidelity}
\label{sec:framework:fidelity}

This dimension evaluates whether the marginal distribution of each
column is faithfully reproduced, using \emph{sample-size adaptive
thresholds} and \emph{multiple complementary metrics}.

For each \textbf{continuous column}, we compute three metrics. The
two-sample KS statistic $D_n = \sup_x |F_{\text{real}}(x) -
F_{\text{synth}}(x)|$ is compared against an adaptive critical value
$\tau_{\text{KS}}(n, \alpha) = c(\alpha) \cdot \sqrt{2/n}$, clamped
to $[0.05, 0.30]$ (see Appendix~\ref{app:details:fidelity} for the
full formula). Jensen--Shannon divergence and total variation distance
are computed over binned distributions, each with threshold 0.20.

For \textbf{categorical columns}, we apply chi-square goodness-of-fit
with cardinality-adjusted significance, plus JSD and TVD over
category frequencies (see Appendix~\ref{app:details:fidelity}).

Scores are aggregated into a table-level fidelity score that
penalizes violations proportionally (Appendix~\ref{app:details:fidelity}).
The dimension passes if the aggregate pass rate exceeds 70\%.
To control the false discovery rate across the many per-column
hypothesis tests, we apply Benjamini--Hochberg (BH) FDR correction
at $q = 0.05$: columns whose KS or chi-square $p$-values are no
longer significant after BH correction are upgraded to ``pass.''
All metrics are formally defined in Appendix~\ref{app:metrics}.

\subsection{Dimension 2: Multivariate Structure}
\label{sec:framework:multivariate}

This dimension evaluates whether inter-column dependencies---the
\emph{joint} distribution, not just marginals---are preserved. Four
sub-metrics provide complementary coverage.

\paragraph{Correlation matrix comparison.}
For tables with $\geq 2$ continuous columns, we compute
$\boldsymbol{\Delta} = |\mathbf{R}_{\text{real}} -
\mathbf{R}_{\text{synth}}|$ (element-wise, diagonal zeroed) using both
Pearson and Spearman methods. The sub-metric passes if
$\Delta_{\max} \leq 0.15$.

\paragraph{Mutual information.}
Normalized mutual information (NMI) captures non-monotonic
dependencies:
$\text{NMI}(X, Y) = I(X;Y) / \sqrt{H(X) \cdot H(Y)}$.
We compare NMI matrices between real and synthetic data; threshold
$\tau_{\text{MI}} = 0.15$.

\paragraph{Contingency tests.}
For categorical column pairs, we compare Cram\'{e}r's V between real
and synthetic contingency tables. The pair passes if
$|V_{\text{real}} - V_{\text{synth}}| < 0.15$.

\paragraph{Conditional distribution checks.}
\label{sec:framework:conditional}
Correlation tests verify the \emph{strength} of dependency but not the
\emph{shape} of $P(Y \mid X)$. We implement two variants. For
continuous $Y$ conditioned on categorical $X$, we compute per-category
KS statistics:
\begin{equation}
  D_v = \text{KS}\bigl(
    Y_{\text{real}} \mid X = v, \;\;
    Y_{\text{synth}} \mid X = v
  \bigr), \qquad
  \bar{D} \leq \tau_{\text{cond}} = 0.25
\end{equation}
For continuous $Y$ conditioned on binned continuous $X$, we partition
$X$ into $B = 4$ quantile bins and compute per-bin KS statistics
analogously.

This check detects a failure invisible to correlation: a generator can
match the overall \texttt{salary} distribution and the global
salary--department correlation while producing identical salary
distributions for engineers (\$120k--\$200k) and administrators
(\$50k--\$90k).

\subsection{Dimension 3: Cross-Table Integrity}
\label{sec:framework:crosstable}

This dimension operates globally, verifying relational structure.

\paragraph{FK referential integrity.}
For each relationship, we count orphaned rows:
$\text{orphans}(T_p, T_c) = |\{r \in \hat{T}_c \mid r[c_c] \notin
\hat{T}_p[c_p]\}|$. Good generators should produce zero orphans.

\paragraph{Cardinality distribution matching.}
For each 1:N relationship, we compare the full distribution of
child counts per parent via a two-sample KS test to detect
\emph{cardinality collapse}. This is shape-agnostic---it catches
differences regardless of whether the real cardinality distribution
is Gaussian, Poisson, or heavy-tailed, unlike a simple mean
comparison.

\paragraph{Cross-table correlation preservation.}
We compute aggregate statistics at the parent level (e.g., mean child
value per parent) and compare the correlation structure between real
and synthetic databases.

\paragraph{Value bounds compliance.}
We count synthetic values outside the $[\min, \max]$ range of real data.

The composite score is:
$s_{\text{cross}} = 0.35 \cdot s_{\text{FK}} + 0.15 \cdot s_{\text{bounds}}
+ 0.15 \cdot s_{\text{cross-corr}} + 0.15 \cdot s_{\text{card}}
+ 0.20 \cdot s_{\text{cond}}$,
where $s_{\text{cond}}$ is the pass rate of cross-table conditional
distribution checks $P(Y_{\text{child}} \mid X_{\text{parent}})$.
FK scores use the violation rate
$s_{\text{FK}} = 1 - \min(1, \text{violations}/\text{total\_child\_rows})$
rather than a fixed denominator.
The dimension passes if $s_{\text{cross}} \geq 0.70$.

\subsection{Dimension 4: ML Utility}
\label{sec:framework:ml}

We adopt the \textbf{TSTR} (Train on Synthetic, Test on Real)
paradigm~\citep{esteban2017real}. An XGBoost model (200 estimators,
\texttt{max\_depth}=6, \texttt{learning\_rate}=0.1) is trained on
synthetic data and evaluated on held-out real data; the efficacy ratio
$\rho = S_S / S_R$ measures how well synthetic data substitutes for
real. Classification uses weighted F1; regression uses $R^2$. The
dimension passes if $\rho \geq 0.90$.

\paragraph{Target column selection.}
The target column is selected by priority: (1)~user-specified via
configuration; (2)~a categorical column with 2--10 classes
(classification); (3)~the first continuous column (regression).
Ideal targets depend on multivariate structure (e.g., \texttt{tier}
predicted from \texttt{income} and \texttt{region}) rather than
near-constant columns.

\subsection{Dimension 5: Privacy}
\label{sec:framework:privacy}

This dimension ensures synthetic data does not memorize real records.
Distances are computed using a \textbf{Gower hybrid metric}:
Hamming distance for categorical columns and normalized Euclidean
for numerical columns, weighted by column-type proportion.

We compute \textbf{DCR} (Distance to Closest Record): the 5th
percentile must exceed $\tau_{\text{DCR}} = 0.10$.
\textbf{NNDR} (Nearest Neighbor Distance Ratio) detects point-copying:
$\text{NNDR}_i = d_1/d_2$, where values near 1 indicate healthy
generation. \textbf{Exact match detection} counts byte-identical rows.
The composite privacy score is:
\begin{equation}
  s_{\text{privacy}} = 0.4 \cdot \min\!\left(1, \frac{\text{DCR}_{5\%}}{2\tau}\right)
  + 0.3 \cdot \min(1, \text{NNDR}_{\text{mean}})
  + 0.3 \cdot \max(0, 1 - 10 \cdot \text{EM}_{\text{unique}}/|\Dsynth|)
\end{equation}

\paragraph{Real-data duplicate normalization.}
Real databases often contain legitimate duplicate records. To avoid
penalizing synthetic rows that match non-unique real records, we
compute real-to-real self-uniqueness and count only exact matches
against \emph{unique} real records ($\text{EM}_{\text{unique}}$).
Both raw and uniqueness-adjusted scores are reported.

\subsection{Dimension 6: Edge-Case Robustness}
\label{sec:framework:edge}

This dimension evaluates production-readiness via seven sub-metrics
(see Appendix~\ref{app:details:edge} for full definitions):

\begin{itemize}[leftmargin=*,nosep]
  \item \textbf{Null pattern preservation}: $|\text{null\_rate}_R -
  \text{null\_rate}_S| \leq 0.05$.
  \item \textbf{Rare category preservation}: $\geq 50\%$ of categories
  with $<1\%$ frequency must appear in synthetic data.
  \item \textbf{Outlier/tail behavior}: relative error at $p_5, p_{95}$
  tails $< 25\%$.
  \item \textbf{Zero-inflation}: zero-rate difference $\leq 0.05$ for
  columns with $> 5\%$ zeros.
  \item \textbf{Mode collapse detection}: flagged if
  $u_{\text{synth}} < 0.3 \cdot u_{\text{real}}$.
  \item \textbf{Value bounds compliance}: count of values outside
  $[\min, \max]$.
  \item \textbf{Percentile accuracy}: $\max_p \epsilon_p \leq 0.20$ at
  $p_5, p_{25}, p_{50}, p_{75}, p_{95}$, where error is IQR-normalized:
  $\epsilon_p = |q_p^R - q_p^S| / (\text{IQR}(R) + \delta)$.
\end{itemize}

The edge-case score is the mean of sub-metric pass rates; the dimension
passes if $s_{\text{edge}} \geq 0.70$.

\section{Experimental Setup}
\label{sec:experiments}

\subsection{Benchmark Datasets}
\label{sec:experiments:dataset}

We evaluate SynEval on two multi-table relational databases of
differing complexity.

\begin{figure}[H]
\centering
\includegraphics[width=0.75\textwidth]{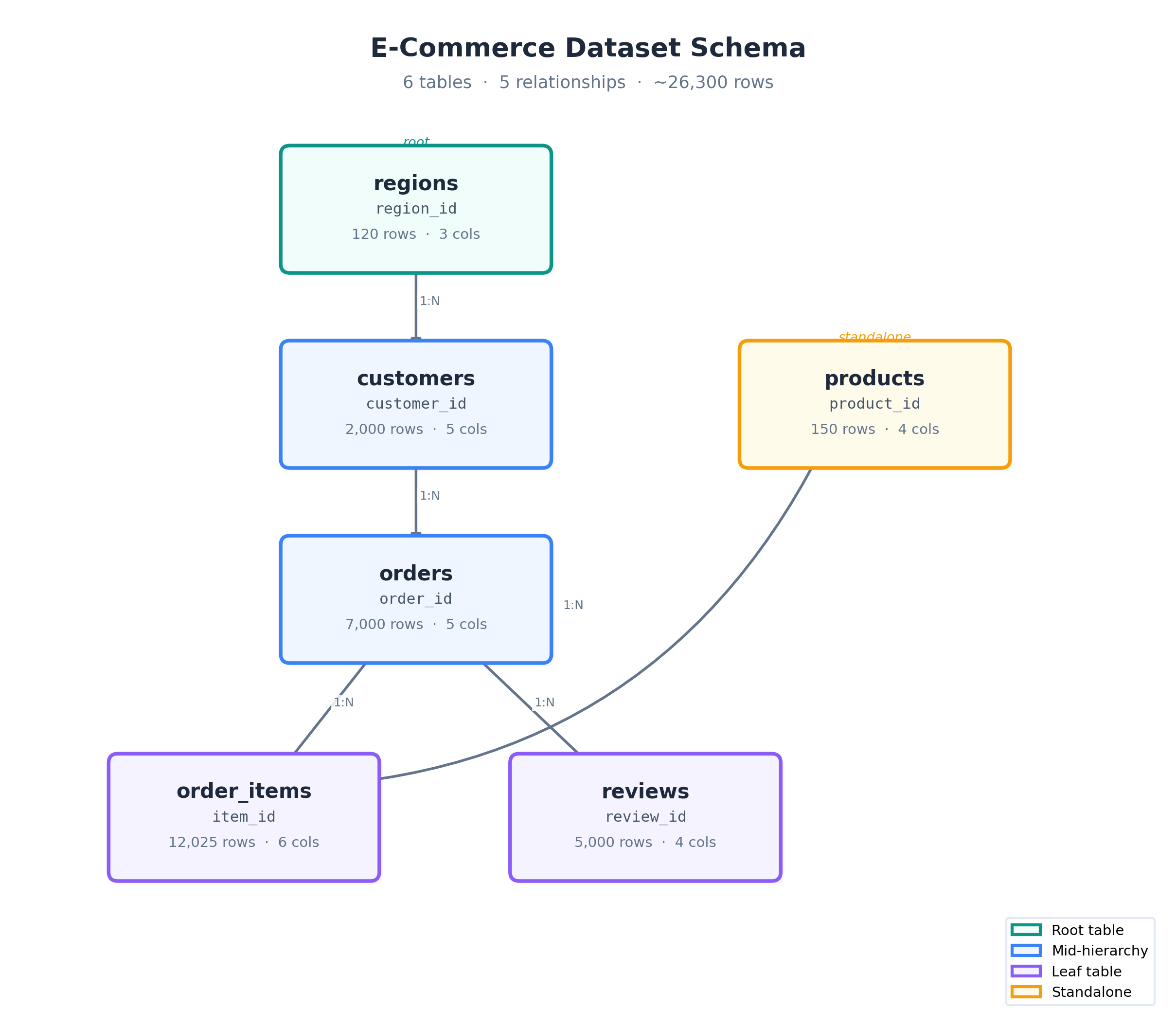}
\caption{E-Commerce dataset schema: entity-relationship diagram
showing 6 tables and 5 parent--child relationships. Color indicates
hierarchy depth: teal~=~root, blue~=~mid-hierarchy,
violet~=~leaf, amber~=~standalone.}
\label{fig:ecom_schema}
\end{figure}

\paragraph{E-Commerce dataset.}
A purpose-built 6-table relational database containing approximately
26{,}300 rows, encoding 13 cross-table edge cases testing categorical-to-continuous parent-child correlations, multi-hop
correlation chains (3--4 hops), conditional variance preservation,
deterministic arithmetic derivations, temporal constraints, exclusive
categorical pairings, and skewed cardinality coupling.
Figure~\ref{fig:ecom_schema} summarizes the schema.

\begin{figure}[H]
\centering
\includegraphics[width=\textwidth]{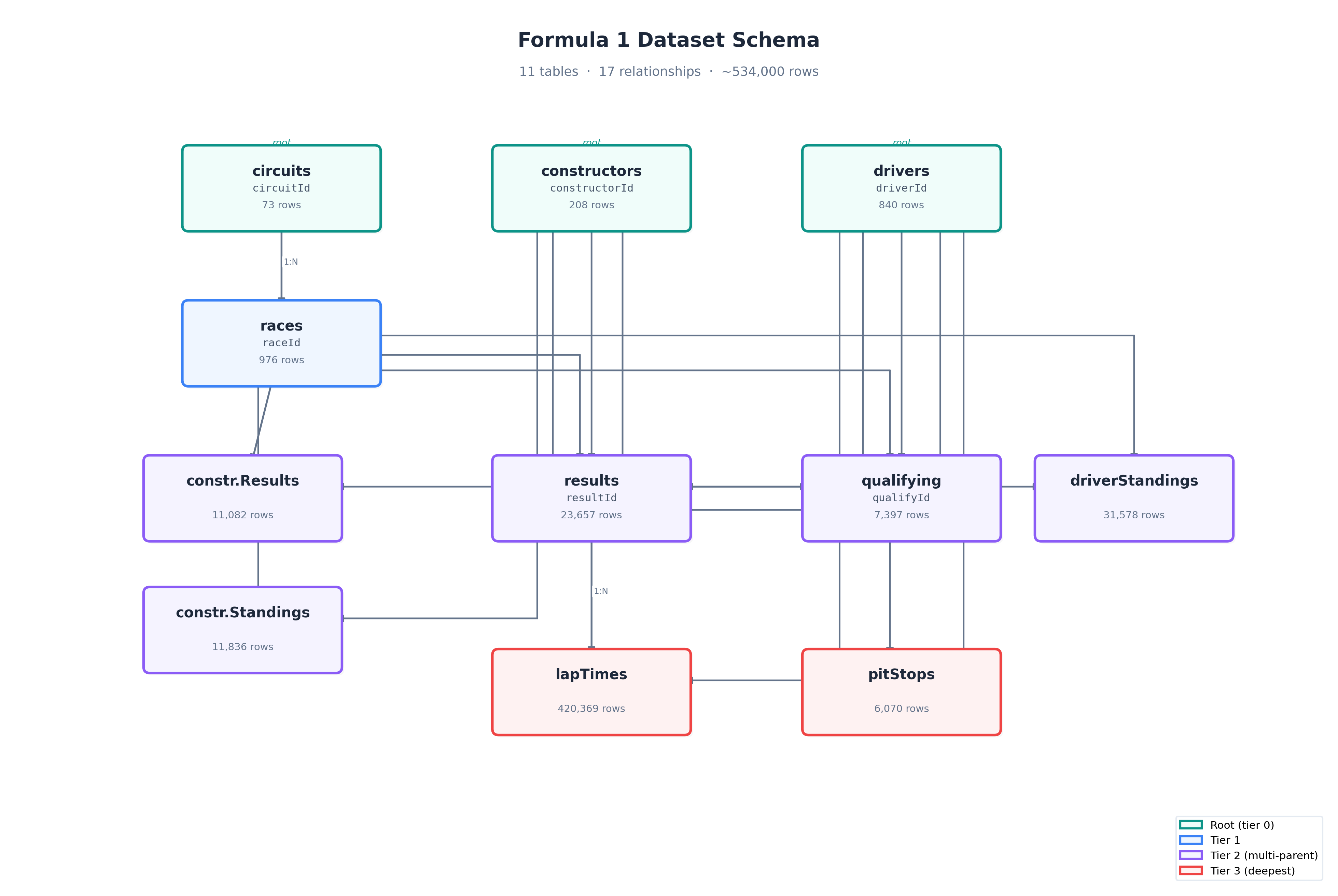}
\caption{F1 dataset schema: entity-relationship diagram showing
11 tables and 17 parent--child relationships. Color indicates
hierarchy depth: teal~=~root, blue~=~tier~1, violet~=~tier~2
(multi-parent), red~=~tier~3 (deepest).}
\label{fig:f1_schema}
\end{figure}

\paragraph{Formula~1 (F1) dataset.}
A real-world 11-table relational database extracted from the Ergast
Developer API, containing approximately 534{,}000 rows across tables
ranging from 73 rows (\texttt{circuits}) to 420{,}369 rows
(\texttt{lapTimes}). The schema features a deep hierarchy
(\texttt{races} $\rightarrow$ \texttt{results} $\rightarrow$
\texttt{lapTimes}) with up to 17 FK relationships and multi-parent
joins (\texttt{results} $\leftarrow$ \texttt{races}, \texttt{drivers},
\texttt{constructors}).
Figure~\ref{fig:f1_schema} summarizes the schema.

The F1 dataset tests SynEval's scalability: it is 20$\times$ larger
than the E-Commerce dataset, has 3$\times$ more relationships, and
contains extreme row-count variation (73 to 420k).

\paragraph{F1 aggregation note.}
The \texttt{lapTimes} table (420{,}369 rows) constitutes 82\% of all
F1 rows, making row-count weighted aggregation disproportionately
dominated by a single table. We therefore use \emph{equal-weight}
averaging across tables for F1. The E-Commerce dataset has more
balanced table sizes and uses the default row-count weighted
aggregation from Equation~(1).

\subsection{Generators Under Evaluation}
\label{sec:experiments:generators}

We evaluate three multi-table generators representing different
architectural paradigms:

\begin{itemize}[leftmargin=*]
  \item \textbf{ClavaDDPM}~\citep{pang2024clavaddpm}: A diffusion-based
  approach using classifier-guided denoising diffusion probabilistic
  models with Gaussian Mixture clustering as latent intermediaries
  between parent and child tables.

  \item \textbf{RC-TGAN}~\citep{gueye2023rctgan}: A GAN-based approach
  that extends conditional tabular GAN training to relational data by
  incorporating parent-row features into the child table generator.

  \item \textbf{RSDG} (Relational Synthetic Data Generator): A
  multi-table extension of conditional tabular GAN training with
  hierarchical parent-to-child sampling and cross-table correlation
  preservation.
\end{itemize}

Each generator was trained with default hyperparameters. No tuning was
performed; the primary goal of this study is to demonstrate SynEval's
evaluation capabilities and diagnostic workflow rather than to conduct
a definitive comparative ranking of generators or datasets. We use
out-of-the-box configurations as representative inputs to illustrate
how the framework surfaces per-dimension strengths and weaknesses.

\subsection{Evaluation Protocol}
\label{sec:experiments:protocol}

SynEval was run on each generator's output with all six dimensions
evaluated using the default weights from Table~\ref{tab:weights}. The
schema was provided via JSON files specifying primary keys, foreign
key relationships, and column type annotations.

\section{Results and Analysis}
\label{sec:results}

\subsection{Overall Results}
\label{sec:results:overall}

Table~\ref{tab:overall} presents the cross-dataset, cross-generator
comparison. ClavaDDPM achieves the highest overall score on both
datasets (87.0\% on F1, 75.9\% on E-Commerce), while on E-Commerce
RSDG matches it closely at 76.0\%.

\begin{table}[!ht]
\centering
\caption{SynEval scores (\%) across datasets and generators.
All six dimensions contribute to the weighted composite.}
\label{tab:overall}
\small
\begin{tabular}{@{}llccc@{}}
\toprule
\textbf{Dataset} & \textbf{Dimension} & \textbf{ClavaDDPM} & \textbf{RC-TGAN} & \textbf{RSDG} \\
\midrule
\multirow{8}{*}{F1}
  & Per-Column Fidelity    & 73.8 & 67.1 & 65.9 \\
  & Multivariate Structure & 83.5 & 73.7 & 76.7 \\
  & Cross-Table Integrity  & 93.8 & 85.8 & 90.3 \\
  & ML Utility             & 26.5 & 56.4 & 58.0 \\
  & Privacy                & 78.3 & 82.1 & 81.0 \\
  & Edge Cases             & 78.0 & 79.1 & 76.8 \\
\cmidrule(lr){2-5}
  & \textbf{Overall (6-dim)}  & \textbf{87.0} & \textbf{80.9} & \textbf{75.2} \\
\midrule
\multirow{8}{*}{E-Com}
  & Per-Column Fidelity    & 70.7 & 67.7 & 64.1 \\
  & Multivariate Structure & 96.6 & 69.1 & 78.8 \\
  & Cross-Table Integrity  & 97.6 & 73.5 & 93.5 \\
  & ML Utility             & 50.0 & 61.3 & 79.5 \\
  & Privacy                & 52.5 & 47.2 & 51.4 \\
  & Edge Cases             & 90.3 & 72.8 & 86.9 \\
\cmidrule(lr){2-5}
  & \textbf{Overall (6-dim)}  & \textbf{75.9} & \textbf{66.7} & \textbf{76.0} \\
\bottomrule
\end{tabular}
\end{table}

\subsection{Radar Comparison}
\label{sec:results:radar}

\begin{figure}[H]
\centering
\includegraphics[width=\textwidth]{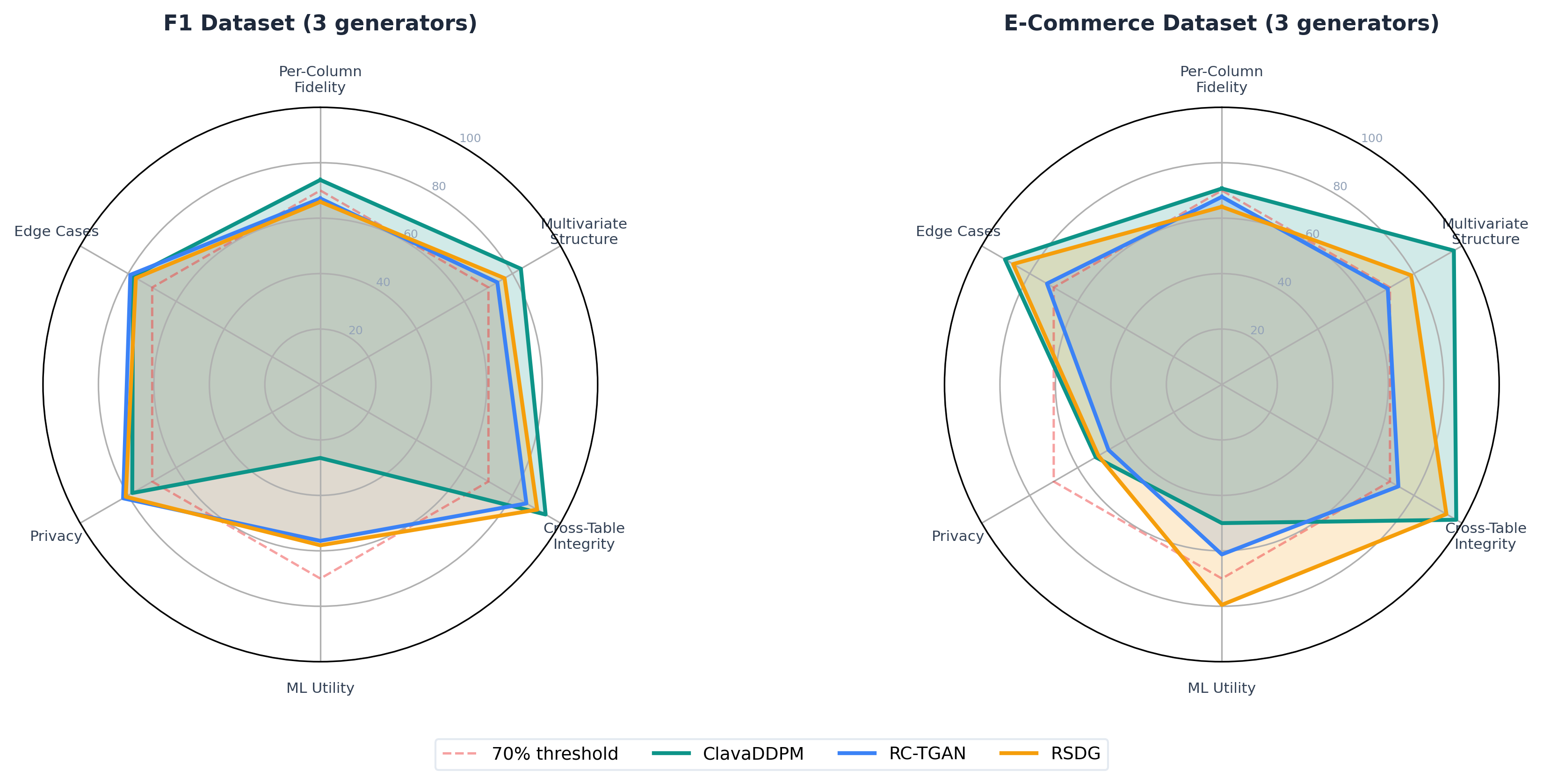}
\caption{Radar charts showing per-dimension scores (6 dimensions).
Left: F1 dataset (3 generators). Right: E-Commerce dataset
(3 generators). The dashed red circle marks the 0.70 pass threshold.}
\label{fig:radar}
\end{figure}

Figure~\ref{fig:radar} shows per-dimension radar plots for both
datasets. ClavaDDPM's polygon consistently encloses the other
generators on structural dimensions (Multivariate, Cross-Table),
while GAN-based generators show stronger ML Utility.

\subsection{ML Utility: A Cross-Dataset Challenge}
\label{sec:results:ml}

ML Utility exhibits the widest variation across generators and
datasets of any dimension, illustrating SynEval's ability to surface
dimension-level disagreements. On F1, ClavaDDPM scores lowest on
ML Utility yet leads on Multivariate Structure and Cross-Table
Integrity; conversely, RSDG and RC-TGAN achieve substantially
higher ML Utility scores while trailing on structural dimensions.
On E-Commerce, RSDG achieves the highest ML Utility and the highest
overall score, while ClavaDDPM leads on structural fidelity but
not on downstream task fitness. These cross-dimension trade-offs---where
strong distributional preservation does not guarantee predictive
utility---are invisible to marginal-only evaluation and demonstrate
the diagnostic value of a multi-dimensional framework.

One contributor to ML Utility variability is target column
sensitivity: the heuristic auto-selection of target columns may
favour columns that are easy or hard to predict depending on schema
structure. Domain-informed target selection---leveraging knowledge
of which columns carry meaningful predictive relationships---would
yield more stable and interpretable ML Utility scores.

\subsection{Privacy: Dataset-Dependent}
\label{sec:results:privacy}

Privacy results are dataset-dependent. On F1, all generators achieve
78--82\%, with the large table sizes (420k rows) making
memorization statistically harder. On E-Commerce, all score below
the pass threshold (47--53\%), with small tables (120--150 rows)
enabling near-memorization. This highlights that privacy risk depends
heavily on dataset size and dimensionality, not just generator
architecture.

\subsection{Cross-Table Integrity}
\label{sec:results:crosstable}

Cross-Table Integrity is consistently the strongest dimension.
ClavaDDPM achieves 93.8\% on F1 and 97.6\% on E-Commerce;
RSDG scores 90.3\% on F1 and 93.5\% on E-Commerce.
RC-TGAN achieves 85.8\% on F1 but drops to 73.5\% on E-Commerce.
All generators' hierarchical parent-to-child sampling structurally
prevents orphaned foreign keys. The score differences stem from
cross-table correlation and conditional distribution preservation
rather than FK violations.

\section{Discussion}
\label{sec:discussion}

\subsection{Multi-Table Synthesis Remains a Hard Problem}

Across six evaluations (3 generators $\times$ 2 datasets),
ClavaDDPM scores highest on F1 at 87.0\% and 75.9\% on E-Commerce.
RC-TGAN achieves 80.9\% on F1 but only 66.7\% on E-Commerce.
RSDG scores 75.2\% on F1 and 76.0\% on E-Commerce---notably
matching ClavaDDPM on the smaller dataset. No generator consistently
exceeds 80\% across both datasets, confirming that multi-table
synthesis with preserved cross-table correlations remains an open
challenge.

\subsection{The Fidelity--Privacy Tension}

Privacy results are dataset-size-dependent. On the large F1 dataset
(534k rows), all generators achieve 78--82\% privacy scores, while on
the small E-Commerce dataset (26k rows), all fail (47--53\%). Small
tables with few distinct values make memorization nearly inevitable
for current generators. This suggests that privacy metrics should
be interpreted relative to dataset size, and future work should
explore size-normalized privacy scores.

\subsection{ML Utility Validates Multi-Dimensional Evaluation}

All generators achieve acceptable marginal fidelity (64--74\%)
yet show wide variation in ML utility (27--80\% across generators
and datasets). A marginal-only evaluation would rate generators
as ``acceptable'' when their synthetic data may be severely degraded for
downstream ML tasks. This disconnect is a strong validation of
SynEval's multi-dimensional diagnostic approach.

\subsection{Experimental Scope}

\begin{enumerate}[leftmargin=*]
  \item Three generators were evaluated with default hyperparameters;
  tuned configurations may yield better results.

  \item The ML Utility dimension depends on target column selection
  and a single model family (XGBoost); different models or targets
  may produce different rankings.

  \item Two datasets were used; results may differ on datasets with
  different schema complexity or domain characteristics.
\end{enumerate}

\section{Conclusion and Future Work}
\label{sec:conclusion}

We have presented SynEval, a six-dimensional evaluation framework for
multi-table synthetic data addressing five critical gaps: marginal-only
metrics, neglect of cross-table integrity, fragmented utility/privacy
assessment, absent production-readiness checks, and lack of unified
scoring. SynEval produces a decomposable quality score across
per-column fidelity, multivariate structure, cross-table integrity, ML
utility, privacy, and edge-case robustness.

The primary goal of this work is to introduce and validate the
evaluation framework itself, not to produce a definitive ranking of
generators. The experimental results presented in
\S\ref{sec:results} demonstrate how SynEval surfaces
dimension-level trade-offs---such as the tension between structural
fidelity and ML utility---that are invisible to marginal-only
assessment. Generator-specific conclusions should be drawn with
caution: in-depth hyperparameter tuning, architecture-specific
data preprocessing, and domain-informed target column selection
may substantially change individual generator scores.

Three evaluation capabilities are, to our knowledge, novel: (1)~conditional
distribution checks $P(Y \mid X \in \text{bin})$ detecting failures
invisible to both marginal tests and correlation comparisons;
(2)~cross-table cardinality distribution matching catching ``cardinality
collapse''; and (3)~edge-case robustness testing covering null patterns,
rare categories, zero-inflation, tail behavior, and mode collapse.

\subsection{Framework Limitations}

\begin{enumerate}[leftmargin=*]
  \item \textbf{Temporal structure not evaluated.} The framework does
  not assess temporal patterns, seasonality, or sequential dependencies.

  \item \textbf{Privacy metrics are distance-based, not differentially
  private.} DCR and NNDR measure empirical distance but do not provide
  formal $(\varepsilon, \delta)$-DP guarantees, requiring access to the
  generation mechanism incompatible with generator-agnostic design.

  \item \textbf{Dimension weights are heuristic.} The default weights
  reflect our assessment of relative importance but are not empirically
  derived.

  \item \textbf{Default training parameters.} All generators in our
  experiments were trained with default hyperparameters; systematic
  hyperparameter tuning during model training may change the relative
  rankings and absolute scores reported here.
\end{enumerate}

\subsection{Future Work}

\begin{enumerate}[leftmargin=*]
  \item \textbf{Temporal evaluation dimension} for seasonal
  decomposition, autocorrelation matching, and event-sequence fidelity.

  \item \textbf{Broader benchmarking} with additional generators
  (SDV/HMA, REaLTabFormer) on multiple datasets of varying schema
  complexity, including healthcare and financial domains.

  \item \textbf{Hyperparameter-guided optimization} using SynEval
  scores as a feedback signal for generator tuning.

  \item \textbf{Privacy-preserving generator evaluation.} Evaluating
  DP-enabled generators (e.g., DPCTGAN) would validate the privacy
  metrics and quantify the fidelity--privacy tradeoff under formal
  $(\varepsilon, \delta)$-DP guarantees.

  \item \textbf{Membership inference attacks.} Augmenting the privacy
  dimension with shadow-model MIA~\citep{shokri2017membership,
  ward2026mtmia} would detect subtler leakage beyond exact-match
  and distance-based metrics.
\end{enumerate}

\section*{Acknowledgements}

We would like to express our deepest gratitude to Robert Sim (rsim@microsoft.com) for his
valuable inputs, guidance, and support throughout this work.


\appendix

\section{Complete Metric Summary}
\label{app:metrics}

Table~\ref{tab:metrics_full} provides a complete reference of all
metrics, their definitions, default thresholds, and pass/fail criteria.

\begin{table}[!ht]
\centering
\caption{Complete metric reference for SynEval.}
\label{tab:metrics_full}
\small
\begin{tabular}{@{}p{2.2cm}p{3.0cm}p{4.5cm}cc@{}}
\toprule
\textbf{Dimension} & \textbf{Metric} & \textbf{Definition} & \textbf{Threshold} & \textbf{Pass if} \\
\midrule
\multirow{4}{*}{\parbox{2.2cm}{Per-Column\\Fidelity}}
  & KS statistic & $\sup_x |F_R(x) - F_S(x)|$ & Adaptive & $\leq \tau$ \\
  & JSD & $\frac{1}{2}D_{KL}(P\|M) + \frac{1}{2}D_{KL}(Q\|M)$ & 0.20 & $\leq 0.20$ \\
  & TVD & $\frac{1}{2}\sum|P(x)-Q(x)|$ & 0.20 & $\leq 0.20$ \\
  & Chi-square & Goodness-of-fit $p$-value & Adaptive & $p > \tau$ \\
\midrule
\multirow{5}{*}{\parbox{2.2cm}{Multivariate\\Structure}}
  & Pearson $\Delta$ & $\max_{i<j}|R_{ij}-\hat{R}_{ij}|$ & 0.15 & $\leq 0.15$ \\
  & Spearman $\Delta$ & Same, rank correlation & 0.15 & $\leq 0.15$ \\
  & NMI $\Delta$ & $\max|NMI_R - NMI_S|$ & 0.15 & $\leq 0.15$ \\
  & Cram\'er's V $\Delta$ & $|V_R - V_S|$ per pair & 0.15 & $< 0.15$ \\
  & Conditional KS & $\text{mean}_b \, D_b$ & 0.25 & $\leq 0.25$ \\
\midrule
\multirow{4}{*}{\parbox{2.2cm}{Cross-Table\\Integrity}}
  & FK violations & Orphaned child rows & 0 & $= 0$ \\
  & Bounds violations & Values outside $[\min,\max]$ & 0 & $= 0$ \\
  & Cross-table corr & Aggregate correlation diff & --- & Score-based \\
  & Cardinality dist & Child-count distribution match & --- & Score-based \\
\midrule
\multirow{2}{*}{\parbox{2.2cm}{ML Utility}}
  & TSTR (classif.) & $\rho = F1_S / F1_R$ (XGBoost) & 0.90 & $\geq 0.90$ \\
  & TSTR (regress.) & $\rho = R^2_S / R^2_R$ (XGBoost) & 0.90 & $\geq 0.90$ \\
\midrule
\multirow{3}{*}{\parbox{2.2cm}{Privacy}}
  & DCR $5^{\text{th}}\%$ & $P_5(\min_j \|\hat{r}_i - r_j\|)$ & 0.10 & $\geq 0.10$ \\
  & NNDR mean & $\text{mean}(d_1/d_2)$ & --- & $\to 1.0$ \\
  & Exact matches & $|\hat{r} \in D_R|$ & 0 & $= 0$ \\
\midrule
\multirow{7}{*}{\parbox{2.2cm}{Edge Cases}}
  & Null rate $\Delta$ & $|\text{null}_R - \text{null}_S|$ & 0.05 & $\leq 0.05$ \\
  & Rare cat.\ pres. & Fraction preserved & 0.50 & $\geq 0.50$ \\
  & Tail error & Rel.\ error at $p_5, p_{95}$ & 0.25 & $<0.25$ \\
  & Zero-rate $\Delta$ & $|\text{zero}_R - \text{zero}_S|$ & 0.05 & $\leq 0.05$ \\
  & Mode collapse & $u_S < 0.3 \cdot u_R$ & --- & Not flagged \\
  & Value bounds & Out-of-range count & 0 & $= 0$ \\
  & Percentile error & $\max_p \epsilon_p$ & 0.20 & $\leq 0.20$ \\
\bottomrule
\end{tabular}
\end{table}

\section{Implementation Details}
\label{app:details}

\subsection{Schema Auto-Inference}
\label{app:details:schema}

When no schema YAML is provided, SynEval auto-infers column types:
object/category dtypes and low-cardinality numerics ($\leq 20$ unique
values) are classified as categorical; remaining numerics as continuous;
columns ending in \texttt{\_id} are excluded as keys. Primary keys are
inferred as the first column with 100\% uniqueness and an \texttt{\_id}
suffix.

\subsection{Per-Column Fidelity Details}
\label{app:details:fidelity}

\paragraph{Adaptive KS threshold.}
The critical value is:
\begin{equation}
  \tau_{\text{KS}}(n, \alpha) =
  c(\alpha) \cdot \sqrt{\frac{2}{n}}, \qquad
  c(\alpha) = \begin{cases}
    1.63 & \alpha \leq 0.01 \\
    1.36 & \alpha \leq 0.05 \\
    1.22 & \text{otherwise}
  \end{cases}
\end{equation}
clamped to $[0.05, 0.30]$.

\paragraph{JSD computation.}
Real and synthetic values are binned into $B = 50$ quantile-based
bins (edges at combined-data percentiles) with Laplace smoothing
($\epsilon = 10^{-10}$). Quantile-based binning adapts to skewed
distributions, unlike equal-width bins that waste resolution on
empty ranges.

\paragraph{Categorical chi-square adjustment.}
The p-value threshold is adjusted by
$\alpha / \max(1, |\mathcal{V}|/5)$ to control the family-wise error
rate for high-cardinality categoricals.

\paragraph{Table-level aggregation.}
\begin{equation}
  s_{\text{fidelity}} = \frac{1}{|\mathcal{M}|}
  \sum_{r \in \mathcal{M}}
  \left(1 - \frac{1}{2} \min\!\left(1, \frac{v_r}{\tau_r}\right)\right)
\end{equation}
where $v_r$ is the metric value and $\tau_r$ its threshold.

\subsection{Multivariate Structure Details}
\label{app:details:multivariate}

\paragraph{Correlation matrix diagnostics.}
In addition to $\Delta_{\max}$, we report mean drift
$\Delta_{\text{mean}} = \frac{2}{p(p-1)} \sum_{i<j} \Delta_{ij}$,
Frobenius norm $\|\boldsymbol{\Delta}\|_F$, and the top 10
most-drifted pairs.

\paragraph{MI discretization.}
Continuous columns are discretized into 10 quantile bins using
\texttt{KBinsDiscretizer}. Pairs where both NMI values are $< 0.01$
(near-independence) are excluded.

\paragraph{Contingency test limits.}
Evaluation is limited to the top $\ell = 15$ categorical pairs to
control computational cost. The pair passes if
$|V_{\text{real}} - V_{\text{synth}}| < 0.15$ and chi-square
$p \geq 0.05$.

\paragraph{Dimension-level aggregation.}
\begin{equation}
  s_{\text{multi}} = \frac{1}{|\mathcal{S}|} \sum_{m \in \mathcal{S}}
  \sigma_m
\end{equation}
where $\sigma_m = \max(0, 1 - \Delta_{\max} / (2 \tau_m))$ for
continuous scoring on Pearson, Spearman, and MI sub-metrics
($\tau_m$ is the respective threshold).

\subsection{Edge-Case Robustness Details}
\label{app:details:edge}

\paragraph{Null pattern preservation.}
$\Delta_{\text{null}}(c) = |\text{null\_rate}_R(c) -
\text{null\_rate}_S(c)| \leq 0.05$.

\paragraph{Rare category preservation.}
Categories with frequency $< 1\%$ form the rare set $\mathcal{R}$.
Preservation rate $= |\mathcal{R} \cap \mathcal{R}_S| / |\mathcal{R}|
\geq 0.50$.

\paragraph{Outlier/tail behavior.}
Outlier rate: $P(|X_c - \mu_c| > 3\sigma_c)$. Tail error:
$|p_R - p_S| / (|p_R| + \epsilon)$ at $p_5$ and $p_{95}$; passes if
$< 0.25$.

\paragraph{Zero-inflation.}
For columns with real zero rate $> 5\%$:
$|\text{zero\_rate}_R - \text{zero\_rate}_S| \leq 0.05$.

\paragraph{Mode collapse.}
Flagged if $u_S / u_R < 0.3$ where $u = |\text{unique}(\hat{T})| /
|\hat{T}|$.

\paragraph{Value bounds.}
$v_c = |\{r[c] < \min T[c]\}| + |\{r[c] > \max T[c]\}|$; any
$v_c > 0$ fails.

\paragraph{Percentile accuracy.}
$\epsilon_p(c) = |q_p^R(c) - q_p^S(c)| /
(\text{IQR}(R_c) + \delta)$ at
$p_5, p_{25}, p_{50}, p_{75}, p_{95}$; passes if $\max_p \epsilon_p
\leq 0.20$. IQR normalization avoids blow-up near zero-valued
percentiles.

\paragraph{Dimension score.}
$s_{\text{edge}} = \frac{1}{|\mathcal{E}|} \sum_{e \in \mathcal{E}}
\text{pass\_rate}(e)$ where $\mathcal{E}$ spans all seven sub-metrics.

\end{document}